\documentclass[11pt,floatfix,notitlepage,byrevtex]{article}
\usepackage{graphicx,epstopdf}
\usepackage[totalwidth=500pt, totalheight=650pt]{geometry}
\usepackage{pstricks,pst-plot,pst-grad,pst-3dplot,pstricks-add,pst-node}
\usepackage{mathrsfs}   %mathscr{},...
\usepackage{amsmath}    %equation*,...
\usepackage{amssymb}    %mathfrank,hslash...
\usepackage{CJKutf8,CJKnumb}
\usepackage{hyperref}
\usepackage{CJK}
\usepackage{amsfonts}
\usepackage{caption2,graphics,placeins,verbatim}
\usepackage{diagbox}
\usepackage[numbers,sort&compress]{natbib}
\usepackage{authblk}

\newcommand{\rmnum}[1]{\romannumeral #1}
\allowdisplaybreaks

\title{Effective field theory on a finite boundary of the Bruhat-Tits tree}
\author{Feng Qu\thanks{qufeng@syu.edu.cn}}
\affil{\small{The Normal College, Shenyang University, Shenyang, P.~R.~China}}
\date{}

\begin{document}
%\begin{CJK}{UTF8}{gbsn}
\maketitle

\begin{abstract}
Based on bulk reconstruction from the finite boundary of the Bruhat-Tits tree, the boundary effective theory is obtained after integrating out fields outside this boundary. According to the $~p$-adic version of Anti-de Sitter/Conformal Field Theory duality, two-point functions of dual theory living on the finite boundary are read out from the effective action. They can be regarded as two-point functions of a deformed conformal field theory over $~p$-adic numbers.
\end{abstract}

\section{Introduction}

It is proposed that physics should be invariant under the change of number fields~\cite{Volovich:1987wu}. For example, we should be able to use either real numbers($~\mathbb{R}~$) or $~p$-adic numbers($~\mathbb{Q}_p~$)~\cite{vvs1988gf,Brekke:1993gf,Vladimirov:1994wi} to set up spacetime coordinates and write down the same physical laws. Such number fields should include the set of rational numbers($~\mathbb{Q}~$) since all measurement resluts in physics are rational numbers. Considering that $~\mathbb{Q}_p~$ and $~\mathbb{R}~$ are the only two candidates satisfying certain restrictions such as including $~\mathbb{Q}~$, it is necessary to study physics over $~\mathbb{Q}_p~$ as investigations to the above proposal. Another motivation to study physics over $~\mathbb{Q}_p~$ comes from the possibility that spacetime is non-Archimidean at small scales~\cite{Volovich:1987wu,Volovich:1987nq,Varadarajan2004}, and it is very convenient to construct such spacetime using $~\mathbb{Q}_p~$. String theories over $~\mathbb{Q}_p~$($~p$-adic string) begin with~\cite{Volovich:1987nq,Freund:1987kt,Freund:1987ck}, and the Bruhat-Tits tree($~\textrm{T}_p~$) is regarded as the $~p$-adic string world-sheet in~\cite{Zabrodin:1988ep}. Spinors, gravity and blackholes on $~\textrm{T}_p~$ are studied in~\cite{Gubser:2018cha,Gubser:2016htz,Huang:2019pgr,Huang:2020qik,Heydeman:2016ldy,Ebert:2019src}. Relations between $~\textrm{T}_p~$ and tensor network are studied in~\cite{Hung:2019zsk,Heydeman:2018qty,Bhattacharyya:2017aly}. The $~p$-adic version of the Anti-de Sitter/Conformal Field Theory duality~\cite{Maldacena:1997re,Gubser:1998bc,Witten:1998qj} is proposed in~\cite{Gubser:2016guj,Heydeman:2016ldy}($~p$-adic AdS/CFT), which are followed by lots of works, such as~\cite{Gubser:2017tsi,Dutta:2017bja,Qu:2018ned,Marcolli:2018ohd,Jepsen:2018dqp,Hung:2018mcn,Jepsen:2018ajn,Qu:2019tyi,Chen:2021rsy,Chen:2021qah,Chen:2021ipv}.

Among all these references,~\cite{Zabrodin:1988ep} and~\cite{Gubser:2016guj} are the most important to this paper. Besides identifying $~\textrm{T}_p~$ as the $~p$-adic string world-sheet,~\cite{Zabrodin:1988ep} also calculates the effective field theory on the infinite boundary of $~\textrm{T}_p~$ which is obtained by integrating out  fields in the bulk. ``Effective'' comes from the integration of fields. One key technique is the use of bulk-boundary propagators. But propagators seem useless when one wants to calculate the effective field theory on the finite(cutoff) boundary, where bulk reconstruction from the finite boundary is required. ``Cutoff'' usually means ignoring one side of the boundary. ``Finite boundary'' is preferred to ``cutoff boundary'' in this paper because both sides of the boundary are handled carefully, and none of them is dropped directly. 

One motivation of this paper is to extend~\cite{Zabrodin:1988ep}'s work which is to calculate the effective field theory on a finite boundary. Bulk reconstruction from the finite boundary is solved in section~\ref{reofbulk}, and the effective field theory is calculated in section~\ref{ourres}. Another motivation is to find some results of $~p$-adic AdS/CFT which are parallel to those of AdS/CFT over $~\mathbb{R}~$ with a cutoff AdS boundary, such as~\cite{Balasubramanian_1999,Boer_2000,McGough:2016lol,Giribet:2017imm,Kraus:2018xrn}. Identifying $~\textrm{T}_p~$ as the $~p$-adic version of AdS spacetime~\cite{Gubser:2016guj}, two-point functions of a deformed CFT over $~\mathbb{Q}_p~$ are calculated in section~\ref{sectwopoint}, where the deformation comes from the ``cutoff'' of $~\textrm{T}_p~$, or in other words, comes from the finite boundary. Section~\ref{btfs} provides some basic knowledge of $~\textrm{T}_p~$ and points out the field space used in this paper. The last section is summary and discussion. In this paper, the measure $~\mu~,~dx~$, the $~p$-adic absolute value $~|\cdot|_p~$ and the edge length $~L~$ have the dimension of length while $~p$-adic numbers are dimensionless.

\section{\label{btfs}The Bruhat-Tits tree and field spaces}

\begin{figure}
	\centering
	\includegraphics[width=0.8\textwidth]{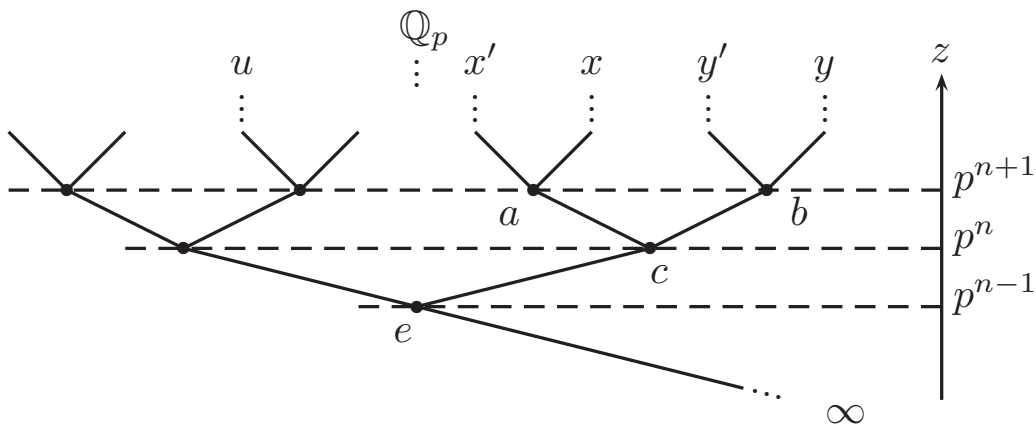}
	\caption{\label{tp}$\textrm{T}_{p=2}~$ and its vertical coordinate $~z~$.}
\end{figure}

Referring to FIG.~\ref{tp},
$\textrm{T}_p~$ is an infinite tree with $~p+1~$ edges incident on each vertex, where $~p~$ is a prime number. Distance $~d(\cdot,\cdot)~$ between vertices can be defined as the number of edges between them. Letting $~z(\cdot)~$ denote the vertical coordinate of a vertex, there is a particular one-to-one correspondence between the upper boundary of $~\textrm{T}_p~$ and $~\mathbb{Q}_p~$ such that $~|x-y|_p=|z(a_{xy})|_p~$, where $~a_{xy}~$ is the lowest vertex on the line connecting $~x~$ and $~y~$ on the upper boundary($~x,y\in\mathbb{Q}_p~$). $~|x-y|_p~$ also defines the distance between $~x~$ and $~y~$, and it is actually the regularization of $~d(a,b)~$ when $~a\to x~$ and $~b\to y~$. Referring to FIG.~\ref{tp}, $~a,b\to x,y~$ can be achieved by $~x'\to x~$ and $y'\to y$. According to~\cite{Gubser:2016guj}, we can write
\begin{gather}
p^{-d(a,b)}=\Big|\frac{(x-x')(y-y')}{(x-y)(x'-y')}\Big|_p\overset{x'(y')\to x(y)}{\sim}\frac{1}{|x-y|_p^2}~,
\end{gather}
where the right-hand side of ``$\sim$'' is the regularization of the left-hand side. There is only one single point on the lower boundary of $~\textrm{T}_p~$, which is noted as $~\infty~$. Each vertex can be regarded as a subset(ball) of $~\mathbb{Q}_p~$ containing points on the upper boundary which are connected to this vertex from above. There is an additive measure $~\mu~$ of vertex $~a~$ which equals to $~|z(a)|_p~$. Several examples are provided in FIG.~\ref{tp}, such as
\begin{gather}
d(a,c)=d(b,c)=d(e,c)=1~,
\\
d(a,e)=d(b,e)=d(a,b)=2~,
\\
|x-y|_p=|z(a_{xy})|_p=|z(c)|_p=|p^n|_p=p^{-n}~,
\\
|x-u|_p=|y-u|_p=|z(e)|_p=|p^{n-1}|_p=p^{1-n}~,
\\
x\in a~,~y\in b~,~u\in e~,~a\cup b=c\subset e~,
\\
p^2\mu(a)=p^2\mu(b)=p\mu(c)=\mu(e)=|z(e)|_p=p^{1-n}~.
\end{gather}
Be aware that $u\notin c$ since edge $~ec~$ is attached to $c$ from below but not from above.

Consider the action and equation of motion of a real-valued massless scalar field on $~\textrm{T}_p~$:
\begin{gather}
S=\frac{1}{2}\sum_{\langle ab\rangle}\frac{(\phi_a-\phi_b)^2}{L^2}~,
\\
\Box\phi_a=0~,~\Box f_a:=\sum_{b\in\partial a}(f_a-f_b)~,
\end{gather}
where $~\langle ab\rangle~$ is the edge connecting the neighboring vertices $~a~$ and $~b~$. The constant $L~$ is the length of edges. $b\in\partial a~$ means $~b~$ is a neighboring vertex of $~a~$ and the sum $~\sum_{b\in\partial a}~$ is over all the neighboring vertices of $~a~$. This action can be rewritten as a sum over vertices, which is
\begin{gather}
4L^2S=\sum_{a}\sum_{b\in\partial a}(\phi_a-\phi_b)^2=2\sum_{z(a)\leq p^N}\phi_a\Box \phi_a+F_N(\phi,\phi)+R_N(\phi,\phi)~,
\\
F_N(f,g):=\sum_{z(a)=p^N}\sum_{\substack{b\in\partial a\\z(b)=p^{N+1}}}(f_a+f_b)(g_b-g_a)~,~R_N(f,g):=\sum_{z(a)>p^N}\sum_{b\in\partial a}(f_a-f_b)(g_a-g_b)~.
\end{gather}
$~R_N~$ comes from the separation
\begin{gather}
\sum_{a}=\sum_{z(a)\leq p^N}+\sum_{z(a)>p^N}
\end{gather}
and $~F_N~$ comes from the identity
\begin{gather}\label{id}
\sum_{z(a)\leq p^N}\sum_{b\in\partial a}(f_a-f_b)(g_a-g_b)=2\sum_{z(a)\leq p^N}f_a\Box g_a+F_N(f,g)=2\sum_{z(a)\leq p^N}g_a\Box f_a+F_N(g,f)~.
\end{gather}
It is convenient to consider a field space where $~R_N~$ and $~F_N~$ vanish. For the field space $~\mathcal{H}~$ in this paper, we demand that
\begin{gather}
\forall f,g\in\mathcal{H}~,~\lim_{N\to\infty}F_N(f,g)=\lim_{N\to\infty}R_N(f,g)=0~.
\end{gather}
Hence, we can always write
\begin{gather}
S=\frac{1}{2}\sum_{\langle ab\rangle}\frac{(\phi_a-\phi_b)^2}{L^2}=\frac{1}{2L^2}\sum_a\phi_a\Box\phi_a~,
\end{gather}
where no boundary term appears.

\section{\label{reofbulk}Bulk reconstruction from the finite boundary}

\begin{figure}
	\centering
	\includegraphics[width=0.8\textwidth]{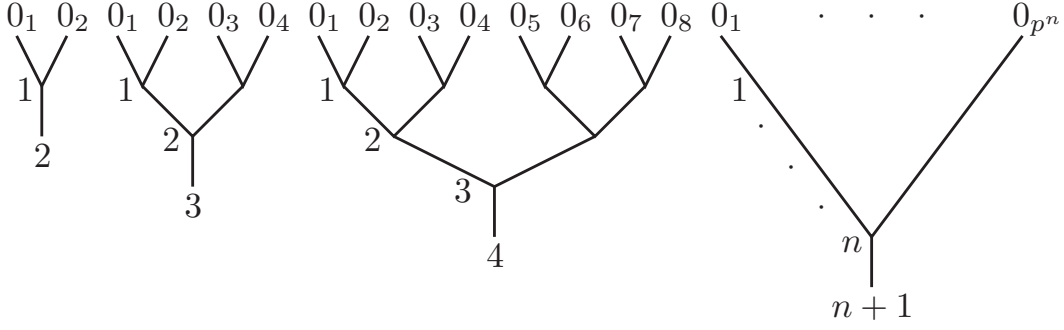}
	\caption{\label{npl1ton}Four subgraphs of $~\textrm{T}_{p=2}~$. Vertices on the upper boundary are noted as $~0_i~$'s. The lower boundary of each subgraph contains only one vertex.}
\end{figure}

With the help of on-shell conditions in the bulk, fields there can be reconstructed from those on the boundary. In FIG.~\ref{npl1ton},
there are four subgraphs of $~\textrm{T}_{p=2}~$. From left to right their bulks and boundaries(bdy) are
\begin{gather}
\begin{aligned}
\textrm{bulk}:\{1\}~,&~\textrm{bdy}:\{0_1,0_2,2\}~,
\\
\textrm{bulk}:\{1,2,\cdots\}~,&~\textrm{bdy}:\{0_1,0_2,0_3,0_4,3\}~,
\\
\textrm{bulk}:\{1,2,3,\cdots\}~,&~\textrm{bdy}:\{0_1,0_2,0_3,0_4,0_5,0_6,0_7,0_8,4\}~,
\\
\textrm{bulk}:\{1,\cdots,n,\cdots\}~,&~\textrm{bdy}:\{0_1,\cdots,0_{p^n},n+1\}~.
\end{aligned}
\end{gather}
Refer to the first subgraph on the right. $\phi_n~$, whose location is one edge above the lower boundary $~\{n+1\}~$, can be reconstructed from $~\phi_{n+1}~$(the field on the lower boundary) and $~\phi_{0}~$'s(fields on the upper boundary). After solving the cases of $~n=1,2,3~$(three subgraphs on the left), the following ansatz can be proposed:
\begin{gather}
a_{n+1}\phi_n=a_{n}\phi_{n+1}+a_1\sum_n\phi_0~,~n\geq1~,\label{ans1}
\\
a_n:=p^n-1~,~\sum_n\phi_0:=\sum_{i=1}^{p^n}\phi_{0_i}~.
\end{gather}
It can be proved by mathematical induction. 

What is useful in this paper is the reconstruction of $~\phi_1~$, whose location is one edge below the upper boundary. Letting $n=1,2,3$ in~(\ref{ans1}), we have
\begin{gather}
\left.
\begin{aligned}
a_2\phi_1=&a_1\phi_2+a_1\sum_1\phi_0
\\
a_3\phi_2=&a_2\phi_3+a_1\sum_2\phi_0
\\
a_4\phi_3=&a_3\phi_4+a_1\sum_3\phi_0
\end{aligned}
\right\}\Rightarrow
\\
\phi_1=\frac{a_1^2}{a_1a_2}\sum_1\phi_0+\frac{a_1^2}{a_2a_3}\sum_2\phi_0+\frac{a_1^2}{a_3a_4}\sum_3\phi_0+\frac{a_1^2}{a_1a_4}\phi_4~.\label{re1forneq3}
\end{gather}
Referring to the third subgraph on the left in FIG.~\ref{npl1ton},~(\ref{re1forneq3}) is the reconstruction of $~\phi_1~$ from $~\phi_{4}~$ and $~\phi_{0}~$'s. Therefore, the ansatz for the reconstruction of $~\phi_1~$ from $~\phi_{n+1}~$ and $~\phi_{0}~$'s can be proposed as
\begin{gather}
\frac{1}{a_1^2}\phi_1=\sum_{i=1}^{n}\frac{1}{a_ia_{i+1}}\sum_i\phi_0+\frac{1}{a_1a_{n+1}}\phi_{n+1}~,
\end{gather}
which can also be proved by mathematical induction. 

Let's consider a simple case of the boundary condition on the lower boundary, which is $~\phi_{a\to\infty}=0~$. Remember that $~\infty~$ is the lower boundary of $~\textrm{T}_p~$(FIG.~\ref{tp}). Letting $~\phi_{n+1}=0~$ and $~n\to\infty~$, the reconstruction of $~\phi_1~$ writes
\begin{gather}\label{infto1}
\frac{1}{a_1^2}\phi_1=\sum_{i=1}^{\infty}\frac{1}{a_ia_{i+1}}\sum_i\phi_0~.
\end{gather}
It can be rearranged into a more useful form. Taking the third subgraph on the left in FIG.~\ref{npl1ton} as an example, we can write
\begin{gather}
\sum_3\phi_0=\sum_{i=1}^{2^3}\phi_{0_i}=\sum_{i=1}^{2^1}\phi_{0_i}+\sum_{i=2^1+1}^{2^2}\phi_{0_i}+\sum_{i=2^2+1}^{2^3}\phi_{0_i}\equiv\sum_{1}\phi_{0}+\sum_{2\setminus1}\phi_{0}+\sum_{3\setminus2}\phi_{0}~,
\end{gather}
where ``$\equiv$'' means that we introduce new symbols on the right-hand side to denote the left-hand side. Remembering that each vertex is a ball in $~\mathbb{Q}_p~$, $\sum_{(i+1)\setminus i}~$ means the sum is over all vertices $~0~$'s(vertices on the upper boundary) included in vertex $~i+1~$ but not included in vertex $~i~$. It can be found that there are $~p~$ terms in $~\sum_1~$ and $~p^{i+1}-p^i~$($~i\geq1~$) terms in $~\sum_{(i+1)\setminus i}~$. Now the reconstruction of $~\phi_1~$~(\ref{infto1}) can be rewritten as
\begin{gather}
\begin{aligned}\label{rewrit}
\frac{1}{a_1^2}\phi_1=&\frac{1}{a_1a_2}\sum_1\phi_0+\frac{1}{a_2a_3}\sum_2\phi_0+\frac{1}{a_3a_4}\sum_3\phi_0+\frac{1}{a_4a_5}\sum_4\phi_0+\cdots
\\
=&\frac{1}{a_1a_2}\sum_1\phi_0+\frac{1}{a_2a_3}(\sum_1\phi_0+\sum_{2\setminus1}\phi_0)+\frac{1}{a_3a_4}(\sum_1\phi_0+\sum_{2\setminus1}\phi_0+\sum_{3\setminus2}\phi_0)
\\
+&\frac{1}{a_4a_5}(\sum_1\phi_0+\sum_{2\setminus1}\phi_0+\sum_{3\setminus2}\phi_0+\sum_{4\setminus3}\phi_0)+\cdots
\\
=&A_1\sum_1\phi_0+A_2\sum_{2\setminus1}\phi_0+A_3\sum_{3\setminus2}\phi_0+A_4\sum_{4\setminus3}\phi_0+\cdots~,
\end{aligned}
\\
A_k=\sum_{i=k}^{\infty}\frac{1}{a_ia_{i+1}}~,~k\geq1~.
\end{gather}
The distance between any vertex $~0_j\subset(i+1)\setminus i~$ and vertex $~1~$ is a constant which only depends on $~i~$. Taking the third subgraph on the left in FIG.~\ref{npl1ton} as an example, we have
\begin{gather}
\begin{aligned}
0_1\cup0_2=1~,&~d(0_1,1)=d(0_2,1)=1=2*1-1~,
\\
0_3\cup0_4=2\setminus1~,&~d(0_3,1)=d(0_4,1)=3=2*2-1~,
\\
0_5\cup0_6\cup0_7\cup0_8=3\setminus2~,&~d(0_5,1)=d(0_6,1)=d(0_7,1)=d(0_8,1)=5=2*3-1~.
\end{aligned}
\end{gather}
Therefore, under the boundary condition $~\phi_{a\to\infty}=0~$, the reconstruction of $~\phi_1~$ from $~\phi_{0}~$'s~(\ref{rewrit}) also writes
\begin{gather}\label{recon1}
\frac{1}{a_1^2}\phi_1=\sum_{n=1}^{\infty}A_n\sum_{d(1,0)=2n-1}^{\infty}\phi_0\equiv\sum_{0\in\textrm{bdy}}A_{\frac{1+d(1,0)}{2}}\phi_0~,
\end{gather}
where $~\sum_{d(1,0)=2n-1}~$ means the sum is over vertices on the upper boundary which are $~2n-1~$ edges away from vertex $~1~$. $~\sum_n\sum_{d(1,0)}\equiv\sum_{0\in\textrm{bdy}}~$ is the sum over all vertices on the upper boundary. The weight coefficient $~A_{(1+d)/2}~$ only depends on the distance between $~\phi_0~$'s location and vertex $~1~$. 

\section{\label{ourres}The effective field theory on the finite boundary}

\begin{figure}
	\centering
	\includegraphics[width=0.8\textwidth]{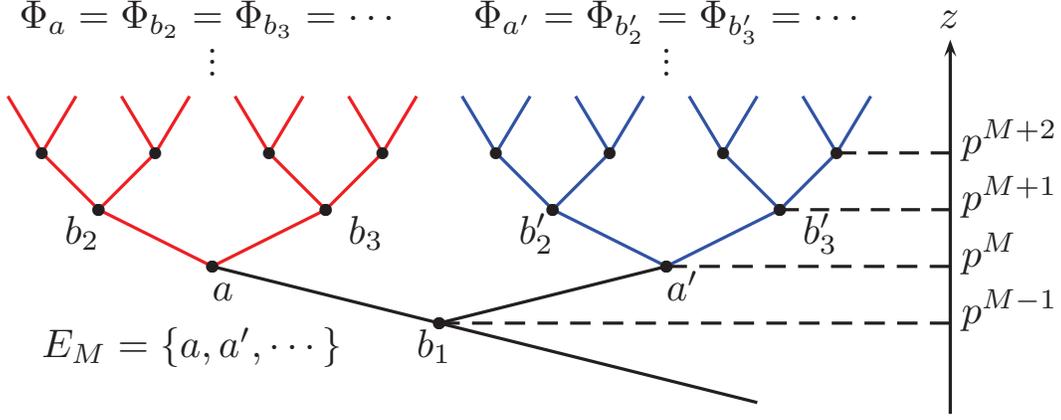}
	\caption{\label{abovecon}The configuration of $~\Phi_a~$ when $~z(a)>p^M~$. Take $~a\in E_M~$ as an example. $~b_1\equiv a^-,b_2~$ and $~b_3~$ are $~p+1=3~$ neighboring vertices of $~a~$, which satisfy $z(b_1)=z(a^-)<z(a)$ and $~z(b_2)=z(b_3)>z(a)~$. $\Phi~$'s on vertices included in $~a~$(vertices of the red subgraph) equal to $~\Phi_a~$. $\Phi~$'s on vertices of the blue subgraph equal to $~\Phi_{a'}~$, and so on.}
\end{figure}

Consider the partition function with sources only living on a finite boundary $~E_M~$. We can write
\begin{gather}
Z_M[J]=\frac{\int_{\textrm{T}_p}\mathcal{D}\phi e^{-S+\sum_{a\in E_M}\phi_aJ_a}}{\int_{\textrm{T}_p}\mathcal{D}\phi e^{-S}}~,
\\
S=\frac{1}{2}\sum_{\langle ab\rangle}\frac{(\phi_a-\phi_b)^2}{L^2}=\frac{1}{2L^2}\sum_a\phi_a\Box\phi_a~,
\\
E_M:=\{a|z(a)=p^M\}~,~J_{a\notin E_M}=0~,
\end{gather}
where $~\int_{\textrm{T}_p}\mathcal{D}\phi~$ means $~\phi~$ fluctuates on the entire $~\textrm{T}_p~$. Decompose $~\phi~$ into $~\Phi~$ and $~\phi'~$ which satisfy
\begin{gather}\label{decompose}
\phi_a=\Phi_a+\phi_a'~,~\Box\Phi_{a\notin E_M}=0~,~\phi_{a\in E_M}'=0~.
\end{gather} 
$~\Phi~$ is on-shell outside $~E_M~$ and $~\phi'~$ vanishes on $~E_M~$. It can be found that $~\Phi~$ and $~\phi'~$ are decoupled in our free field theory, and only $~\Phi~$ will contribute to the final result. Rewriting the action using $~\Phi~$ and $~\phi'~$, we have
\begin{gather}
2L^2S=\sum_{a\in E_M}\Phi_a\Box\Phi_a+S'=\sum_{a\in E_M}\Phi_a(\Phi_a-\Phi_{a^-})+\sum_{a\in E_M}\Phi_a(p\Phi_a-\sum_{\substack{b\in\partial a\\z(b)>z(a)}}\Phi_b)+S'~,
\\
S'=\sum_a\phi_a'\Box\phi_a'~,
\end{gather}
where~(\ref{id}) and~(\ref{decompose}) are used. Among $~p+1~$ neighboring vertices of $~a~$, there is only one satisfying $~z(b)<z(a)~$(noted as $~a^-~$) and the rest satisfying $~z(b)>z(a)~$. When choosing a particular on-shell configuration of $~\Phi_a~$ above $~E_M~$($~z(a)>p^M~$), the second term in the action vanishes, and it makes the calculation easier. Referring to FIG.~\ref{abovecon},
we have
\begin{gather}\label{decay}
p\Phi_a-\sum_{\substack{b\in\partial a\\z(b)>z(a)}}\Phi_b=p\Phi_a-\sum_{\substack{b\in\partial a\\z(b)>z(a)}}\Phi_a=0~.
\end{gather}
Other on-shell configurations which are not considered in this paper, such as $~\Phi_b=p^{-1}\Phi_a~$ in~(\ref{decay}), can introduce a non-zero mass term. According to the reconstruction of $~\phi_1~$ from $~\phi_0~$'s~(\ref{recon1}), $\Phi_{a^-}~$ can be reconstructed from $~\Phi~$'s on $~E_M~$. And the action can be written as
\begin{gather}
2L^2S=\sum_{a\in E_M}\Phi_a(\Phi_a-\Phi_{a^-})+S'=\sum_{a\in E_M}\Phi_a(\Phi_a-a_1^2\sum_{b\in E_M}A_{\frac{1+d(a^-,b)}{2}}\Phi_b)+S'~.
\end{gather}
Considering that there are $~p~$ vertices($~b~$'s) satisfying $~d(a^-,b)=1~$ and $~p^{n}-p^{n-1}~$ vertices satisfying $~d(a^-,b)=2n-1~$ when $~n\geq2~$, it can be proved that
\begin{gather}
\begin{aligned}
a_1^2\sum_{b\in E_M}A_{\frac{1+d(a^-,b)}{2}}=&a_1^2\Big (A_1p+A_2(p^2-p)+A_3(p^3-p^2)+\cdots\Big)
\\
=&a_1^2\Big(p(A_1-A_2)+p^2(A_2-A_3)+p^3(A_3-A_4)+\cdots\Big)
\\
=&a_1(\frac{a_2-a_1}{a_1a_2}+\frac{a_3-a_2}{a_2a_3}+\frac{a_4-a_3}{a_3a_4}+\cdots)=1~.
\end{aligned}
\end{gather}
Hence the action also writes
\begin{gather}
2L^2S=a_1^2\sum_{a\in E_M}\Phi_a\Big(\sum_{b\in E_M}A_{\frac{1+d(a^-,b)}{2}}(\Phi_a-\Phi_b)\Big)+S'=a_1^2\sum_{a\in E_M}\Phi_a\Big(\sum_{\substack{b\in E_M\\b\neq a}}A_{\frac{d(a,b)}{2}}(\Phi_a-\Phi_b)\Big)+S'~.
\end{gather}
Substituting it into the partition function, terms related to $~\phi'~$ cancel out. And it turns out to be a partition function of a field theory on $~E_M~$, which is
\begin{gather}
Z_M[J]=\frac{\int_{E_M}\mathcal{D}\Phi e^{-S_M+\sum_{a\in E_M}\Phi_aJ_a}}{\int_{E_M}\mathcal{D}\Phi e^{-S_M}}~,
\\
S_M=\frac{(p-1)^2}{2L^2}\sum_{a\in E_M}\Phi_a\Big(\sum_{\substack{b\in E_M\\b\neq a}}A_{\frac{d(a,b)}{2}}(\Phi_a-\Phi_b)\Big)~.\label{effonfinite}
\end{gather}
$\int_{E_M}\mathcal{D}\Phi~$ means $~\Phi~$ only fluctuates on $~E_M~$, which comes from the separation
\begin{gather}
\int_{\textrm{T}_p}\mathcal{D}\phi=\int_{E_M}\mathcal{D}\phi\int_{\textrm{T}_p\setminus E_M}\mathcal{D}\phi~.
\end{gather} 

\begin{figure}
	\centering
	\includegraphics[width=0.8\textwidth]{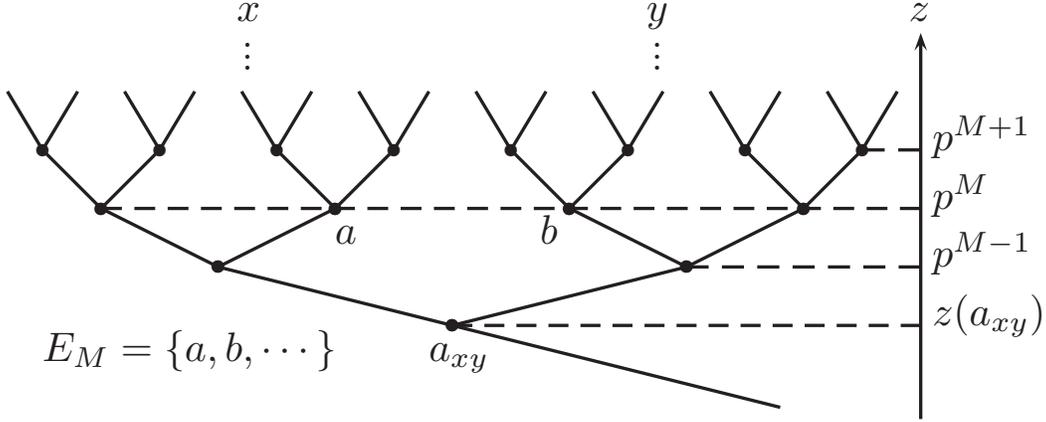}
	\caption{\label{gotoinf}The relation between $~d(a,b)~$ and $~|x-y|_p=|z(a_{xy})|_p~$ where $~x\in a~,~y\in b~$. $d(a,b)~$ is the number of edges between $~a~$ and $~b~$. $~a_{xy}~$ is the lowest vertex on the line connecting $~x~$ and $~y~$. It can be found that $\log_pp^M-\log_pz(a_{xy})=\frac{1}{2}d(a,b)~$, which also writes $~z(a_{xy})=p^{M-d(a,b)/2}~$. In this figure, we have $~d(a,b)=4~,~z(a_{xy})=p^{M-2}~$.}
\end{figure}

(\ref{effonfinite}) is the effective field theory on the finite boundary $E_M$. Taking the limit $~M\to\infty~$ leads to that on the infinite boundary. Refer to FIG.~\ref{gotoinf}. 
Given vertices $~a,b\in E_M~$, select two points $~x~$ and $~y~$ on the upper boundary of $~\textrm{T}_p~$ satisfying $~x\in a~,~y\in b~$. We can write
\begin{gather}\label{fixxy}
|x-y|_p=|z(a_{xy})|_p=|p^{M-\frac{d(a,b)}{2}}|_p=p^{\frac{d(a,b)}{2}}|p^M|_p~.
\end{gather} 
The action $~S_M~$ can be rewritten as
\begin{gather}
\begin{aligned}
\frac{2L^2}{(p-1)^2}S_M=&\sum_{a\in E_M}|p^{M}|_p\Phi_a\Big(\sum_{\substack{b\in E_M\\b\neq a}}|p^{M}|_p\frac{A_{\frac{d(a,b)}{2}}}{|p^M|_p^2}(\Phi_a-\Phi_b)\Big)
\\
=&\sum_{a\in E_M}|p^{M}|_p\Phi_a\Big(\sum_{\substack{b\in E_M\\b\neq a}}|p^{M}|_pA_{\frac{d(a,b)}{2}}p^{d(a,b)}\frac{\Phi_a-\Phi_b}{|x-y|_p^2}\Big)~,
\end{aligned}
\end{gather} 
where $~x\in a~,~y\in b~$. $~|p^{M}|_p~$ is the measure of each vertex on the finite boundary $~E_M~$, which tends to $~dx~$ in the limit $~M\to\infty~$. Supposing that $~a\to x~$ and $~b\to y~$ when $~M\to\infty~$, we can write $~\Phi_a\to\Phi_x~$ and $~\Phi_b\to\Phi_y~$ where $~\Phi_x~$ or $~\Phi_y~$ represents a field on the upper boundary(infinite boundary) of $~\textrm{T}_p~$. As for the $~A_{d/2}p^d~$ term, considering that $~M\to\infty\Leftrightarrow d(a,b)\to\infty~$ according to~(\ref{fixxy}) when fixing $~x~$ and $~y~$, we can write
\begin{gather}
A_{\frac{d(a,b)}{2}}p^{d(a,b)}=p^d\sum_{i=d/2}^{\infty}\frac{1}{(p^i-1)(p^{i+1}-1)}\xrightarrow[]{M\to\infty}p^d\sum_{i=d/2}^{\infty}\frac{1}{p^ip^{i+1}}=\frac{p}{p^2-1}~.
\end{gather}
Finally, in the limit $~M\to\infty~$, the action $~S_M~$ can be written as
\begin{gather}%\label{smoninf}
S_{M\to\infty}=\frac{p(p-1)}{2(p+1)L^2}\int_{x\in\mathbb{Q}_p}dx\Phi_x\int_{\substack{y\in\mathbb{Q}_p\\y\neq x}}dy\frac{\Phi_x-\Phi_y}{|x-y|_p^2}~.
\end{gather}

\begin{figure}
	\centering
	\includegraphics[width=0.8\textwidth]{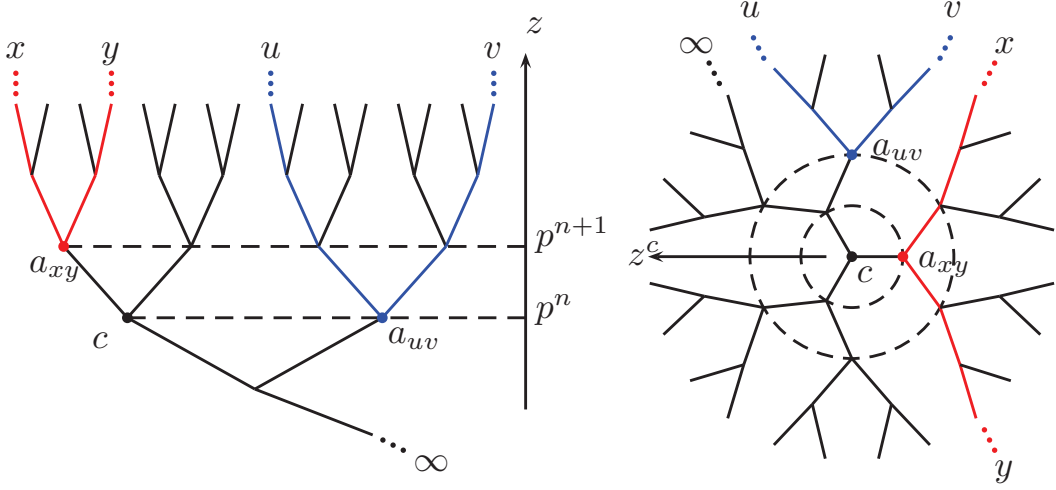}
	\caption{\label{diff}Two different layouts for the same graph $~\textrm{T}_{p=2}~$. Different $~\mu~$'s and $~|\cdot|_p~$'s can be introduced according to different layouts.}
\end{figure}

This effective field theory on the infinite boundary of $~\textrm{T}_p~$ is consistent with~\cite{Zabrodin:1988ep}. But different $~dx~$(or $~\mu~$) and $~|\cdot|_p~$ are used in that paper. The relation between $~dx~$ and $~\mu~$ is $\int_{x\in a}dx=\mu(a)$. Refer to FIG.~\ref{diff}. 
In the left figure, we already know that
\begin{gather}
|x-y|_p=\mu(a_{xy})=|z(a_{xy})|_p=p^{-n-1}~,
\\
|u-v|_p=\mu(a_{uv})=\mu(c)=|z(c)|_p=p^{-n}~.
\end{gather}
The right figure is another layout for the same graph. There is a radial coordinate $~z^c~$ of vertices depending on the distance between this vertex and the reference one $~c~$. For example, we can write
\begin{gather}
z^c(a_{xy})=p^{d(a_{xy},c)}=p^1~,~z^c(a_{uv})=p^{d(a_{uv},c)}=p^2~,~z^c(c)=p^{d(c,c)}=1~.
\end{gather}
Each vertex(noted as $~a~$) in the right figure is a ball in $~\mathbb{Q}_p\cup\{\infty\}~$ containing boundary points which are on the ``half-line'' $~ca~$'s(half-lines which start from $~c~$, pass through $~a~$ and go to the boundary). The reference vertex $~c~$ contains all the boundary points, namely $~c=\mathbb{Q}_p\cup\{\infty\}~$. Measure $\mu^c$ of vertices and distance $~|\cdot|_p^c~$ of boundary points can be introduced according to the right figure, which satisfy
\begin{gather}
|x-y|_p^c=\mu^c(a_{xy})=|z^c(a_{xy})|_p=p^{-1}~,
\\
|u-v|_p^c=\mu^c(a_{uv})=|z^c(a_{uv})|_p=p^{-2}~.
\end{gather} 
They are different from $~\mu~$ and $~|\cdot|_p~$ in the left figure or which are used in this paper. For example, it can be found that
\begin{gather}
\mu(a_{xy})<\mu(a_{uv})~,~\mu^c(a_{xy})>\mu^c(a_{uv})~,
\\
|x-y|_p<|u-v|_p~,~|x-y|_p^c>|u-v|_p^c~.
\end{gather}
$~\mu^c~$ and $~|\cdot|_p^c~$ are the measure and distance used in~\cite{Zabrodin:1988ep}.

\section{\label{sectwopoint}Relations to $~p$-adic AdS/CFT}

Consider the equation
\begin{gather}
Z_M[J]=\frac{\int_{\textrm{T}_p}\mathcal{D}\phi e^{-S+\sum_{a\in E_M}\phi_aJ_a}}{\int_{\textrm{T}_p}\mathcal{D}\phi e^{-S}}=\frac{\int_{E_M}\mathcal{D}\phi\int_{\textrm{T}_p\setminus E_M}\mathcal{D}\phi e^{-S+\sum_{a\in E_M}\phi_aJ_a}}{\int_{E_M}\mathcal{D}\phi\int_{\textrm{T}_p\setminus E_M}\mathcal{D}\phi e^{-S}}=\frac{\int_{E_M}\mathcal{D}\Phi e^{-S_M+\sum_{a\in E_M}\Phi_aJ_a}}{\int_{E_M}\mathcal{D}\Phi e^{-S_M}}~.
\end{gather}
Ignoring denominators and setting $~J=0~$, we can write
\begin{gather}
\int_{\textrm{T}_p\setminus E_M}\mathcal{D}\phi e^{-S}\sim e^{-S_M}\xrightarrow[]{M\to\infty}\int_{\textrm{T}_p}\mathcal{D}\phi e^{-S}\sim e^{-S_{M\to\infty}}~.
\end{gather}
Therefore, $~S_M~$($~S_{M\to\infty}~$) can be regarded as the effective action after integrating out fields on $~\textrm{T}_p\setminus E_M~$($~\textrm{T}_p~$). Now let's identify $~\textrm{T}_p~$ as a $~p$-adic version of AdS spacetime~\cite{Gubser:2016guj}. According to the spirit of AdS/CFT: 
\begin{gather}%\label{adscft}
\langle e^{\int dxO\phi_0}\rangle_{\textrm{CFT}}=\int_{\textrm{AdS}}\mathcal{D}\phi e^{-S}\Big|_{\phi_{\partial\textrm{AdS}}=\phi_0}~,
\end{gather}
where $~\partial\textrm{AdS}~$ is the boundary of AdS and the fluctuation of gravity has been ignored, $~e^{-S_{M\to\infty}}~$ should be directly proportional to the generating functional of some CFT over $~\mathbb{Q}_p~$, whose two-point function reads
\begin{gather}\label{greeninfin}
\frac{\delta^2e^{-S_{M\to\infty}}}{\delta\Phi_x\delta\Phi_{y\neq x}}\Big|_{\Phi=0}=\frac{p(p-1)}{(p+1)L^2}\frac{1}{|x-y|_p^2}~.
\end{gather}
It is consistent with~\cite{Gubser:2016guj} if setting $~\eta_p=1~,~\Delta=n=1~$ there and $~L=1~$ in~(\ref{greeninfin}). On the other hand, if not taking the limit $~M\to\infty~$, the following calculation should give a two-point function of some deformed CFT over (coarse-grained) $~\mathbb{Q}_p~$:
\begin{gather}\label{greenfin}
\frac{\delta^2e^{-S_{M}}}{\delta\Phi_a\delta\Phi_{b\neq a}}\Big|_{\Phi=0}=\frac{(p-1)^2}{L^2}A_{\frac{d(a,b)}{2}}=\frac{(p-1)^2}{L^2}\sum_{n=\frac{d(a,b)}{2}}^{\infty}\frac{1}{(p^n-1)(p^{n+1}-1)}~,
\end{gather}
where $~d(a,b)=2,4,6,8,\cdots~$ is a positive even number and $~E_M=\{a,b,\cdots\}~$ is a coarse-grained $~\mathbb{Q}_p~$. Remember that each element in $~E_M~$ is a ball in $~\mathbb{Q}_p~$. (\ref{greenfin}) can be regarded as a counterpart to the two-point function of a deformed CFT living on the cutoff boundary of AdS over $~\mathbb{R}~$. 

\section{Summary and discussion}

In this paper, we manage to reconstruct fields in the bulk from those on the finite boundary of $~\textrm{T}_p~$~(\ref{recon1}). Then with the help of calculating techniques in~\cite{Zabrodin:1988ep}, the effective field theory is calculated by integrating out fields on the entire $~\textrm{T}_p~$ except those on the finite boundary~(\ref{effonfinite}). According to the spirit of AdS/CFT, two-point functions of dual theories are read out:~(\ref{greeninfin}) on the infinite boundary and~(\ref{greenfin}) on the finite boundary. The former is a two-point function of a CFT over $~\mathbb{Q}_p~$ which is consistent with~\cite{Gubser:2016guj}, and the latter is a two-point function of a deformed CFT which should be compared with that in AdS/CFT over $~\mathbb{R}~$ with a cutoff AdS boundary.

Some problems still need to be explored. For example, \rmnum{1})relations between field spaces discussed in section~\ref{btfs} and those in~\cite{Huang:2019pgr,Qu:2019tyi} are still unclear. Different field spaces or boundary conditions sometimes lead to different results; \rmnum{2})it may be a hard problem to find out what ``deformed CFT'' is which gives a two-point function like~(\ref{greenfin}). It is known that the counterpart over $~\mathbb{R}~$ can be regarded as a $~T\bar{T}$-deformed CFT~\cite{McGough:2016lol}; \rmnum{3})the same calculation on $~p\textrm{AdS}~$ is interesting. $p\textrm{AdS}~$~\cite{Gubser:2016guj,Qu:2018ned} is another $~p$-adic version of AdS spacetime whose finite boundary is exactly $~\mathbb{Q}_p~$ but not the coarse-grained one.

\section*{Acknowledgements}

This work is supported by NSFC grant no.~11875082.

%\end{CJK}
\end{document}